# A New Perspective on Scale: A Novel Transform for NMR Envelope Extraction

Ehsun Assadi

*Abstract*— Envelope extraction in nuclear magnetic resonance (NMR) is a fundamental step for processing the data space generated by this technique. Envelope detection accuracy improves with increasing the number of sampling points; however, we propose a novel transform that enables acceptable envelope extraction with significantly fewer sampling points, even without meeting the Nyquist rate. In this paper, we challenge the traditional scale definition and demonstrate that classic scaling lacks a physical referent in all situations. To achieve this aim, we introduce a scale based on the variations of space-invariant states, rather than the observable characteristics of matter and energy. According to this definition of scale, we distinguished two kinds of observers: scale-variant and scale-invariant. We demonstrated that converting a scale-variant observer to a scale-invariant observer is equivalent to envelop extraction. To analyse and study the theories presented in the paper, we have designed and implemented an Earth-field NMR setup and used real data generated by it to evaluate the performance of the proposed envelope-detection transform. We compared the output of the proposed transform with that of classic and state-of-the-art methods for parameter recovery of NMR signals.

## I. INTRODUCTION

Earth-field nuclear magnetic resonance (EF-NMR) technique utilizes the Earth's magnetic field as the main magnetic field, making it susceptible to noise. Envelope extraction of noisy NMR spin-echo signals is a fundamental step in generating nuclear magnetic resonance images. To achieve higher-frequency estimation precision and accurate envelope detection, a relatively long measurement time and a higher signal-to-noise ratio (SNR) are required. In this paper, we introduce a novel transform that uses more sampling points for spectral analysis than the companion methods, making it a powerful tool for precise frequency estimation at lower frequencies and effective envelope extraction.

In this study, we use Free Induction Decay (FID) and spin-echo NMR signals. FID signals are transient responses from a spin system after an RF pulse excitation (see Fig. 1(a)). Two successive RF pulses produce a spin-echo signal. We designed an Earth-field Nuclear Magnetic Resonance (NMR) system to observe and study spin-echo signals under varying magnetic field homogeneities, which paved the way for the theoretical discussion in the paper (see Fig. 2).

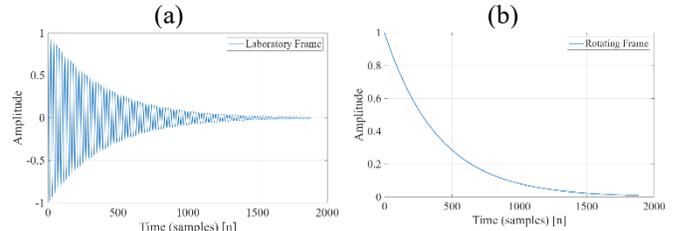

**Fig. 1.** Different frames in NMR technology. (a) Laboratory frame. (b) Rotating frame.

Briefly describe the characteristics of the designed NMR here. A polarizer driver capable of delivering 10A to generate a $25\ mT$ polarization field is implemented and designed, and the RF transmitter and receiver coils are designed for an 8 cm field of view, as shown in Fig. 2(a). An RF transmitter circuit for generating a pulse sequence is implemented and designed (see Fig. 2(b)). The scale simulator for simulating scale and field shimming has been designed and implemented, as shown in Fig. 2(c). A power amplifier with an overall voltage gain of $10^5$ and frequency bandwidth of 300 Hz is implemented to amplify the microvolt signals from the receiver coil, as seen in Fig. 2(d). Two FID and spin-echo signals generated by our setup and arising from a tube of water that is exposed to high field homogeneity and low field homogeneity are shown in Fig. 3(b) and (e), respectively. These signals will be used to study the proposed transform's theory and, finally, to evaluate its performance and that of other methods.

The proposed transform is applied as envelope detection to NMR signals in this paper; therefore, we review previous studies on NMR signal envelope extraction in the following. Envelope detection via the Hilbert transform has been widely adopted in NMR and broader signal processing applications. The Hilbert transform provides an analytic representation of a real-valued signal, from which the instantaneous amplitude (envelope) can be extracted. This technique is particularly effective in isolating slowly varying modulations embedded within oscillatory signals.

Recent works highlight the evolution of surface NMR methodologies from traditional time-domain envelope detection to frequency-domain and steady-state approaches. Grombacher et al. [1] introduced a spectral analysis (SA) envelope detection scheme that leverages discrete Fourier transforms over sliding windows of NMR data to produce high-SNR envelopes.

Liu et al. [2] developed a comprehensive workflow for steady-state free precession (SSFP) surface NMR data, introducing methods such as transient culling, pulse windowing, de-spiking,

Corresponding author: Ehsun Assadi, (e-mail: ehsun.assadi@gmail.com).

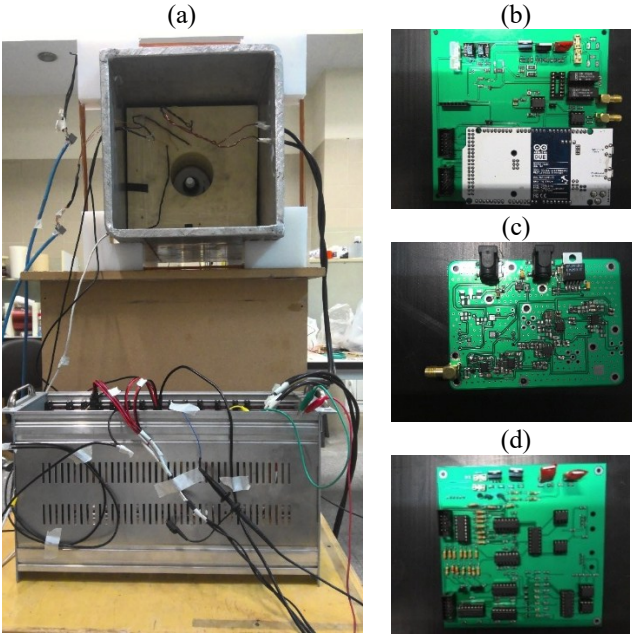

**Fig. 2.** Earth-field NMR setup. (a) Polarizer, transmitter, and receiver coils inside an electromagnetic shield. (b) Transmitter and receiver RF board. (c) Scale simulator. (d) Power amp.

powerline harmonics subtraction, and multiwindow spectral analysis. The narrowband collapse and multiwindow concatenation improve the stability of envelope extraction.

Although these methods present strong contributions to NMR envelope extraction, they trade off envelope continuity and sample richness. A key limitation of these approaches is that the final envelope signals are inherently down-sampled: by relying on sliding-window discrete Fourier transforms centered at specific frequencies, the number of extracted envelope samples is much fewer than the original time-domain signal. This reduction in sampling density can limit temporal resolution and potentially obscure fine-scale variations in the envelope, particularly in early-time data where accurate amplitude tracking is critical.

Tian et al. [3] introduced a novel approach that combines third-order cumulants (TOC) with quadratic frequency coupling (QFC) constraints to enhance the robustness of MRS signal analysis. Their method leverages the property that higher-order cumulants suppress Gaussian random noise, while the QFC constraint selectively retains frequency components that satisfy nonlinear coupling relationships. Their contribution provides a complementary perspective by addressing noise suppression and parameter recovery via nonlinear frequency-coupling constraints.

The theoretical contribution of this article lies in introducing a new perspective on the frames used in NMR technology, which is shown in Fig. 1. In describing the NMR technique, multiple reference frames are used to uniquely define physical quantities, such as the magnetic field and the angular momentum. Two main frames that are constantly referred to in the NMR technique, the laboratory frame and the rotating frame, are shown in Fig. 1(a) and (b), respectively. In the literature, the rotating frame is considered a frame of reference that rotates with respect to the laboratory frame at a specific angular velocity, in both classical and quantum views [4], [5].

In this paper, we introduce a new perspective on scale that helps us redefine the frames in Fig. 1 as observations made by a unique observer, each from a different point of view. This novel perspective enables us to easily transfer between the laboratory and rotating frames, which correspond to the same envelope extraction.

## II. PROBLEM DEFINITION

Regardless of the spectral distribution of a spin system, the maximum amplitude (at t = 0) of the FID signal depends on the magnetic field strength and is reached at t = 0. The decay rate of an FID signal, known as the relaxation time $T_2$, is related to the spectral distribution of magnetized spins $p(w)$.

The spectral density function $p(w)$ determines the characteristics of an FID signal in the laboratory frame. If the receiver coil has a homogeneous reception field over the region of interest, as is often assumed, the FID signal resulting from an $\alpha$ RF pulse takes the following form [5]:

$$S(t) = \sin \alpha \int_{-\infty}^{+\infty} \rho(\omega) e^{-\frac{t}{T_2}} e^{-i\omega t} d\omega \qquad t \geq 0. \quad (1)$$

Suppose that the spectral distribution of a spin system in the laboratory frame can be defined as a spectral distribution with a frequency shift as $\hat{\rho}(\omega) = \rho(\omega + \omega_0)$. Where $\hat{\rho}(\omega)$ is the spectral distribution of magnetized spins in the laboratory frame and $\rho(\omega)$ represents the spectral distribution of magnetized spins in the rotating frame. The FID signal in the laboratory frame becomes:

$$S_{labratory}(t) = \sin \alpha \int_{-\infty}^{+\infty} \rho(\omega) e^{-\frac{t}{T_2}} e^{-i\omega t} d\omega$$
$$= \sin \alpha \int_{-\infty}^{+\infty} [\hat{\rho}(\omega) e^{-i\omega t} d\omega] \, e^{-\frac{t}{T_2}} * e^{i\omega_0 t} \quad (2)$$

The spectral distribution of magnetized spins $\rho(\omega)$ is dependent on the object's inherent properties and is affected by the main magnetic field inhomogeneity. When both the magnetic field and the sample that is exposed to the magnetic field are highly homogeneous, the FID signal exhibits a distinctive $T_2$ decay; otherwise, it decays with a lower time constant named $T_2^*$. FID and spin-echo signals of a tube of water exposed to higher and lower field homogeneity are shown in Fig. 3(b) and (e), respectively.

The question is, what if we consider the dilation of an independent variable, such as time on a timescale, as the leading cause of the different decay rates of the FID signal in Fig. 3(b) and (e)? In other words, we consider the dilation or expansion of the time scale of frames to be the cause of the change in decay rate in the FID signal, rather than changes in spin energy states resulting from different magnetic field homogeneities.

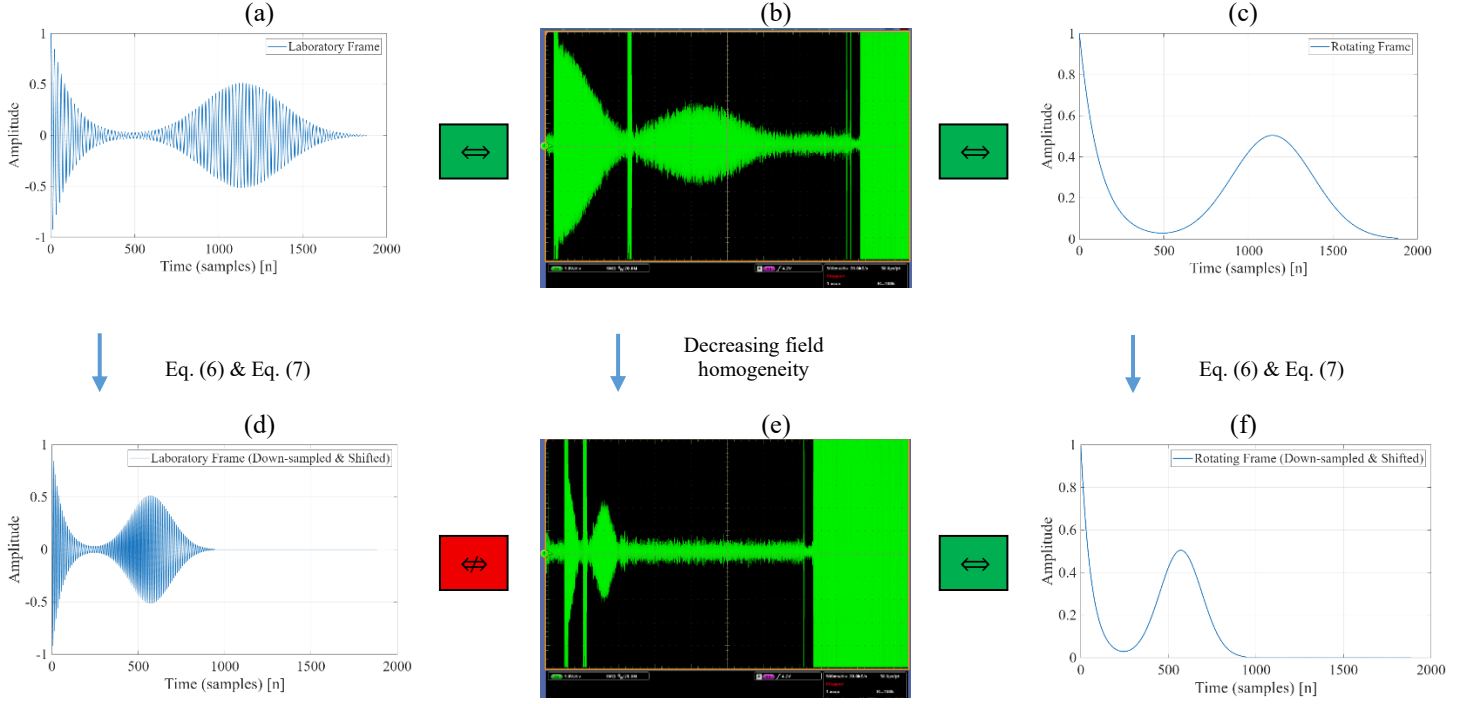

**Fig. 3.** Effect of time-scale scaling on different frames. (a) Synthesized signal in the laboratory frame. (b) FID and spin-echo signal of a tube of water exposed to a high field homogeneity. (c) Synthesized envelope signal in the rotating frame. (d) Synthesized signal in the laboratory frame. (e) FID and spin-echo signal of a tube of water exposed to a low field homogeneity. (f) Synthesized envelope signal in the rotating frame.

It is essential to define the time scale we intend to use in this paper precisely.

The time-scale used in the paper has four main properties:

- First, the uniform motion is a rectilinear motion that is shaped by the motion of a free particle that includes at least three states (not necessarily distinct) of energy during the motion of the free particle.
- Second, the motion of a free particle is uniform on the time scale.
- Third, equal intervals of time are those in which a free particle travels equal distances.
- Fourth, the least interval of time is determined by the observer, not the motion of the free particle.

We must determine whether the time dilation of the synthesized signals in the laboratory frame (Fig. 3(a)), leading to Fig. 3(d), corresponds to the physical observations shown in Fig. 3(b) and Fig. 3(e) within the introduced time scale. According to the properties of the time-scale defined above, when the carrier signal ($e^{i\omega_0 t}$) experiences time-scale scaling, the envelope should exhibit the same kind of scaling. The argument of the carrier signal is linear in time; hence, the envelope's argument should be linear in time as well. The only way to satisfy this is that the field distribution inhomogeneity ($B = B_0 + \Delta B_0$) should lead to a Lorentzian distribution as follows:

$$\rho(w) = M_0 \frac{\Delta\omega_0}{(\Delta\omega_0{}^2 + (\omega - \omega_0)^2)} \qquad (3)$$

Assuming that the magnetic field inhomogeneity with Lorentzian distribution is time-invariant, the FID signal in the rotating frame becomes:

$$S_{labratory}(t) = \pi \sin\alpha\, M_0 * e^{-\frac{t}{T_2}} * e^{-\Delta\omega_0 t} * e^{i\omega_0 t} \qquad (4)$$

$$S_{rotating}(t) = \pi \sin\alpha\, M_0 * e^{-(\frac{1}{T_2} + \Delta\omega_0)t} \qquad (5)$$

Where $M_0$ is the thermal equilibrium value for the bulk magnetization at t=0 in the presence of $B_0$, $\omega_0$ is the Larmor precession angular frequency, $T_2$ is the transverse relaxation time, and $\alpha$ is the angle between the coil's axis and the external magnetic field. The energy difference between the adjacent spin states due to the external magnetic field strength B0, and the total number of spins at room temperature, are considered to remain constant over time-scale changes.

In Fig. 3(a) and (c), we generated two synthetic NMR signals in laboratory and rotating frames, respectively, using Eq. (4)-(5). By increasing $\Delta\omega_0$ in Eq. (4), we simulate the NMR signal exposed to lower-field homogeneity, as shown in Fig. 3(e), and present the results in Fig. 3(d) and 3(f). Overall, the field homogeneity in both synthetic and real signals leads to a Lorentzian distribution.

Time-scale scaling is not directly supported in discrete-time systems, as they have a fixed sampling frequency. Thus, time scaling in discrete-time consists of two parts: down-sampling the signal and shifting the remaining sample points to the scaled time scale with equal intervals but smaller intervals. In discrete-time, downsampling does not change the signal's frequency content; however, the time interval is stretched. To keep the sampling frequency fixed during downsampling in the discrete domain, we shift the downsampled samples to the left.

First, we zero-pad the original signal by a factor of M; second, downsample the zero-padded signal as follows:

$$x_p = \begin{cases} x[n], & 0 \leq n < N - 1 \\ 0, & N \leq n < MN - 1 \end{cases} \qquad (6)$$

$$y[n] = x_p \left[ \frac{n}{M} \right], \qquad 0 \leq n < MN - 1 \qquad (7)$$

Where, $n$ is sampling points, and $M$ is the downsampling factor. After compressing the signal using Eq. (7), we use interpolation to generate new samples with non-integer indices. The above time-scale shrinkage does not violate the optimal sampling pattern for measuring the NMR spin-spin relaxation time [6].

The results of such time scaling in discrete-time do not have a physical referent, see Fig. 3. The time-scale transform results for the rotating frame in Fig. 3(f) are consistent with the physical observation of decreasing field homogeneity, as shown in Fig. 1(e), while the scaled results for the laboratory frame in Fig. 3(d) did not match with the physical observation of decreasing field homogeneity, as shown in Fig. 1(e). Since the frequency of the signal is changed in Fig. 3(c), while the frequency of signals with two different field homogeneities in the real experiments is the same, as shown in Fig. 3(b) and (e).

In continuous time, the mathematical transform we used for time scaling was the Galilean transform, which preserves the laws of electromagnetic fields. Besides, frames are mathematical metric spaces that remain metrics during the transform. It seems the problem is related to the scale at which two observers, one in the laboratory and one in a rotating frame, experience it. We try to define a unified scale for both observers in the frames. To solve the problem, we introduce the scale for time-scaling based on the observer's properties rather than observable characteristics of matter and energy.

First, we need to define the observer we use in the paper. We confine the observer to some constraints for considering scaling in time as an invariant variable to account for the variation of $T_2^*$. This approach enables us to find a physical referent for the time scaling in the discrete time we used above. For having a physical referent for such mathematical time-scaling in a discrete domain, we may confine the observer to three rules:

- First, the observer should have the same frequency sampling rate in all frames we study.
- Second, the observer's observation is time-invariant in all frames.
- Third, the observer must observe at least three samples, in which two of them must be related to distinct states or two different levels of energy. This enables us to define smaller quantized steps, which are required by the fourth property of the time-scale.

Based on the first property of the time-scale and the third property of the observer, no observer observes a mass in the time-scale that travels with a constant speed without changing the amount of mass or energy. This property helps us overcome the classic scale problem, as seen in Fig. 3. To observe such a constant-velocity mass, the observer should sample the rising and falling times of the mass's motion. A mass with a higher speed will have a sharper rising and falling time than a mass with a slower speed. The observer who observes the higher speed mass should have a higher sampling frequency to be able observe such a sharper edge, but this violates the first property of the observer. To maintain the same sampling frequency for both observers, the times of the two observers will differ. In this paper, we intend to use a unique observer with varying points of view; in other words, the observers observe the same mass in the same energy states from different points of view.

Based on the above discussion, the frames in Fig. 1 are not frames; instead, they are observations by the same observer with two different points of view on the state of energy being observed. The different points of view in this paper relate to scale, which we will explain in the following section. Contrary to the frame concept, which holds that different observers with varying sampling frequencies observe masses at different speeds or in different energy states, our concept holds that observers observe the same mass at the same energy level from two different points of view, with a constant sampling frequency. Therefore, we suggest that readers change the word frame in this paper to a point of view to achieve a more precise understanding of the concept the paper aims to introduce. By following these rules for observers, the observer should sample the signals at the same frequency.

Now, we define the scale we used in the paper. Variation in space-invariant states, such as spin, is considered a scale rather than a change in physical size, space, or time intervals. In other words, the frequency of the space-invariant state variations is regarded as a scale in the paper. According to our definition of the scale, there are two types of observers: scale-variant and scale-invariant, both of which meet the three observer properties we introduced.

The difference between a scale-variant and scale-invariant observer is based on their knowledge about either the physical scale or the state of the phenomenon being observed. The observer who observes at least two different space-invariant states of the object out of three samples with varying intervals of time is called a scale-variant observer.

Moreover, there are two types of scale-variant observers: one that observes states continuously and the other that observes them discretely or quantitatively. To avoid burdening the paper, we avoid introducing these two types of scaled observers; instead, we introduce an equivalence principle. We need to introduce an equivalence relation between scale-variant and scale-invariant observers to relate two frames from different points of view. We will use this equivalence in the proposed transform. A scale-invariant observer is equivalent to the discrete scale-variant observer that observes only a unique state.

One of the physical properties to distinguish between a scale-invariant observer and a scale-variant observer is the relation between energy and the frequency of changing the states. The coefficient $\frac{energy}{frequency}$ ($e/f$) is only constant for a scale-invariant observer during the classic scaling procedure. The scale-variant observer will consider that the energy and, consequently, the mass of the object is changed when the scale, in its classic definition, of the time-scale changes. If mathematically scaling the time-scale, as described as down-sampling and shifting in Eq. (6)-(7), one wants to have a physical referent, $^e/_f$ should remain constant during the scaling only for those observers that are free of scale, in its new definition, but have the same sampling frequency in a metric space. However, $^e/_f$ cannot remain constant for a scale-variant observer who considers the same object to have lower energy or lower mass at different scales applied to the time scale.

For instance, the constant $^e/_f$ is fixed by changing field homogeneity in our experiments, as shown in Fig. 3(b) and (e). However, it changes through the classic time-scale scaling (downsampling and shifting process) in the laboratory frame in Fig. 3(b) and (d). The spin energy, powered by the Earth's magnetic field, and the Larmor frequency in high- and low-field homogeneities are the same in Fig. 3(b) and (d), respectively. It means that $^e/_f$ remains constant in both cases; however, the Larmor frequency changes with time scaling in Fig. 3(a) and (d), which means the spins' energy should decrease due to time dilation.

To get back to our main problem, considering the homogeneity effect as a time-scaling effect, we have to change the scale-variant observation, a laboratory view (see Fig. 1 (a)), to a scale-invariant observation as a rotating frame (see Fig. 1 (b)), while both observations experience the same energy. In the following, we introduce a novel transform that converts the scale-variant observer into a scale-invariant observer.

## III. Proposed Transform

In this section, we introduce the proposed transform that converts scale-variant observations into scale-invariant ones. According to the equivalence, a scale-invariant observer must observe a unique state out of many states. Additionally, the observer must observe at least three samples, which include at least two distinct states; otherwise, based on the introduced time-scale properties, changing the energy state of a body will not be distinguishable from a free particle with uniform motion that is scaled in space over time. One of those three samples can depend on the other two, but each of the two samples should relate to two distinct states. To have such an observer, the observer should observe samples in both the present and the future simultaneously. To this end, the signal should be transformed into a complex signal, known as the analytic signal, in which time must be treated as an invariant variable.

Traditional classical approaches to representing and analyzing signals include integral transforms such as the Fourier, Laplace, and Hilbert transforms. The Hilbert transform is a classical approach for converting real signals into analytic signals. The Hilbert transform of a signal is defined as the signal in which all frequency components of the original signal have their phase shifted by 90°. The shifted signal can be interpreted as the signal in the future [7], [8].

The analytic signal is the signal in quadrature with the original signal; when added to the original signal, the oscillating signal becomes a rotating vector. As shown in Fig. 4, an analytic signal is represented by a phasor rotating in the complex plane. In this paper, we discuss the discrete complex space. A phasor can be viewed as a vector at the origin of the complex plane, having a length A(t) and an angle, or an angular position (displacement) $\psi(t)$ [9].

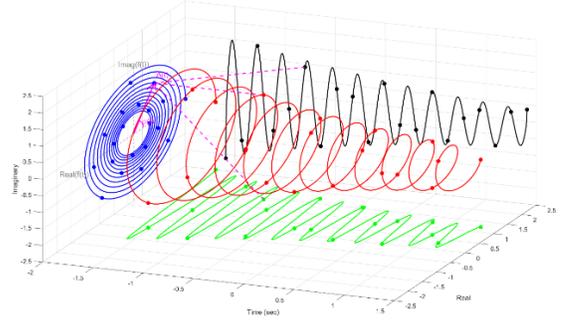

**Fig. 4.** Signal expression in complex domain (the figure inspired by [9]).

As discussed, the introduced scale is independent of mass for a scale-invariant observer. Thus, to change a scale-variant observer to a scale-invariant observer, we need a variable that is independent of mass and energy, and a metric that is a constant, but infinitely divisible quantity. In the complex plane representation, length $A(t)$ and the least spatial interval $\Delta A(t)$ of the Hilbert transform is proportional to mass; therefore, the multiplication of any scale of $\Delta A(t)$ depends on mass and energy. In complex coordinates, the length $A(t)$ is related to mass; however, the least spatial interval in the complex plane is the instantaneous phase $\Delta\psi(t)$ which is independent of mass. Moreover, as shown in Fig. 4, it is a constant metric that can be discretized an infinite number of times. Hence, $\Delta\psi$ is the third observation generated by the two other samples in the real and imaginary coordinate systems, which satisfy the third rule defined for the observation.

The instant frequency $\omega(t)$ measures the rate and direction of a phasor rotation in the complex plane. The instantaneous frequency of the red signal in Fig. 4 for a scale-invariant observer is the same central angular frequency $w_0$ and only the length $A(t)$ changes by time; thus, the mean value of the instantaneous frequency is the same as the central frequency $w_0$. Thus, $w_n = w_0 = \bar{w}$ and $\Delta\psi_n = \Delta\psi$, where $\bar{w}$ is the average angular speed. The least spatial interval instantaneous phase in discrete form can be obtained as follows [9]:

$$w_0 = w_n = \frac{\Delta\psi_n}{\Delta t} = \frac{\psi_{n+1} - \psi_n}{t_{n+1} - t_n} \quad (8)$$

$$\Delta\psi = 2\pi f_0 \Delta t \quad (9)$$

Assuming that the least interval $\Delta t$ is the sampling rate in the discrete form, yielding the final coefficient $\Delta\psi = 2\pi\frac{f_0}{f_s}$. In the proposed transform, we call $f_0$ transfer frequency.

As discussed, $\Delta\psi$ is a variable independent of mass, and a basis generated from it will distinguish rectilinear motion from rotation and acceleration, removing the scale dependence of the scale-variant observer.

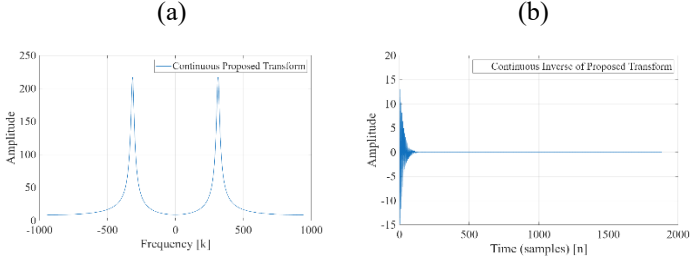

**Fig. 5.** Proposed Transform in continuous form. (a) continuous transform. (b) Abs of inverse continuous transform.

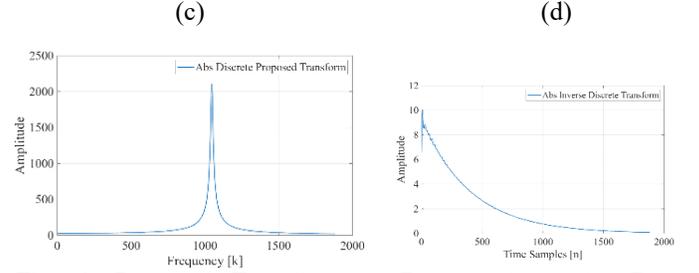

**Fig. 6.** Proposed Transform in Discrete form. (a) Discrete transform. (b) Abs of inverse discrete transform.

The desired analytic signal will be generated by the basis that contains the above coefficient, such that $\frac{f_o}{f_s} < 1$. Thus, we multiply this coefficient by all the sine and cosine basis of the Fourier transform. The proposed method in continuous form for a synthetized FID signal is given by:

Forward continuous transform:

$$F(2\pi f) = \int_{-\infty}^{+\infty} \cos(2\pi f_0 t)\, e^{-\alpha t} e^{-\frac{2\pi f_0 f t}{f_s}}\, dt \qquad (10)$$
$$= \frac{\frac{f_s}{f_o}(j2\pi f + \frac{f_s}{f_o}\alpha)}{(j2\pi f + \frac{f_s}{f_o}\alpha)^2 + (\frac{f_s}{f_o}2\pi f_0)^2}$$

Inverse continuous Transform:

$$F^{-1}(t) = \frac{1}{2\pi}\int_{-\infty}^{+\infty} F\left(2\pi\frac{f_s}{f_o}f\right) e^{\frac{2\pi f_0 f t}{f_s}}\, dt \qquad (11)$$
$$= \frac{f_s}{f_o}\cos(\frac{f_s}{f_o}2\pi f_0 t)\, e^{-\frac{f_s}{f_o}\alpha t}$$

The proposed method in discrete form for a finite signal with finite samples can be expressed as:

Forward discrete transform:

$$X[k] = \sum_{n=0}^{N-1} x[n] e^{-\frac{j2\pi f_0 nk}{Nf_s}}, \quad k \in [0, N-1] \qquad (12)$$

Inverse discrete transform:

$$x[n] = \frac{1}{2\pi N}\sum_{k=0}^{N-1} X[k] e^{\frac{j2\pi f_0 nk}{Nf_s}}, \quad n \in [0, N-1] \qquad (13)$$

$$Envelope = |x[n]| \qquad (14)$$

Where, $f_o$ and $f_s$ are transfer and sampling frequencies, respectively. Also, N is the total number of samples.

The results of applying the forward and inverse continuous transforms to the FID signal in Fig. 1(a) are shown in Fig. 5. The continuous form in Eq. (10) corresponds to the classic time-domain scaling, resulting in scale-variant observations. Its output is the scaled version of the original signal, which lacks a physical referent, as discussed in Section II and shown in Fig. 3. In addition, the continuous inverse form does not produce an analytic, complex signal. The reason is that $\Delta t$, which is considered the inverse of the sample rate, does not hold in continuous form. Because of its infinite frequencies, it does not have a constant sampling frequency throughout the transform, violating the second rule of the observer. The discrete form with a fixed number of samples will solve the problem. The inverse transform of the discrete form generates an analytic signal whose absolute value will change the scale-variant observation to a scale-invariant observation. The output of Eq. (13) is an analytic signal that consists of real and imaginary parts. The absolute value of the analytic signal in Eq. (14) results in an envelope signal. The results of applying the forward and the inverse discrete transform on the FID signal in Fig. 1(a) are shown in Fig. 6.

Thanks to the complex signal generated by the inverse of the proposed transform, its main advantage is that adherence to the Nyquist rate is unnecessary; it suffices for the number of samples to equal the transfer frequency ($f_o$). In other words, we can violate the Nyquist rate in the proposed method, which can be reduced to the transform's transfer frequency, and the sampling frequency and number of samples can be reduced accordingly.

## IV. RESULTS

### A. D-delayed Pulse Study

To better understand the power of the proposed method for frequency estimation precision, we study the rectangular pulse function, see Fig. 7. Three situations for pulse study are considered: a pulse with D-delay in Fig. 7(a), a pulse with 2D-delay in Fig. 7(b), and a pulse with D-delay while the transfer frequency in Eq. (12) is double of the transfer frequency used for the former pulses. All pulses have a fixed pulse width and are sampled with the same sampling frequency equal to $100\ samples/s$.

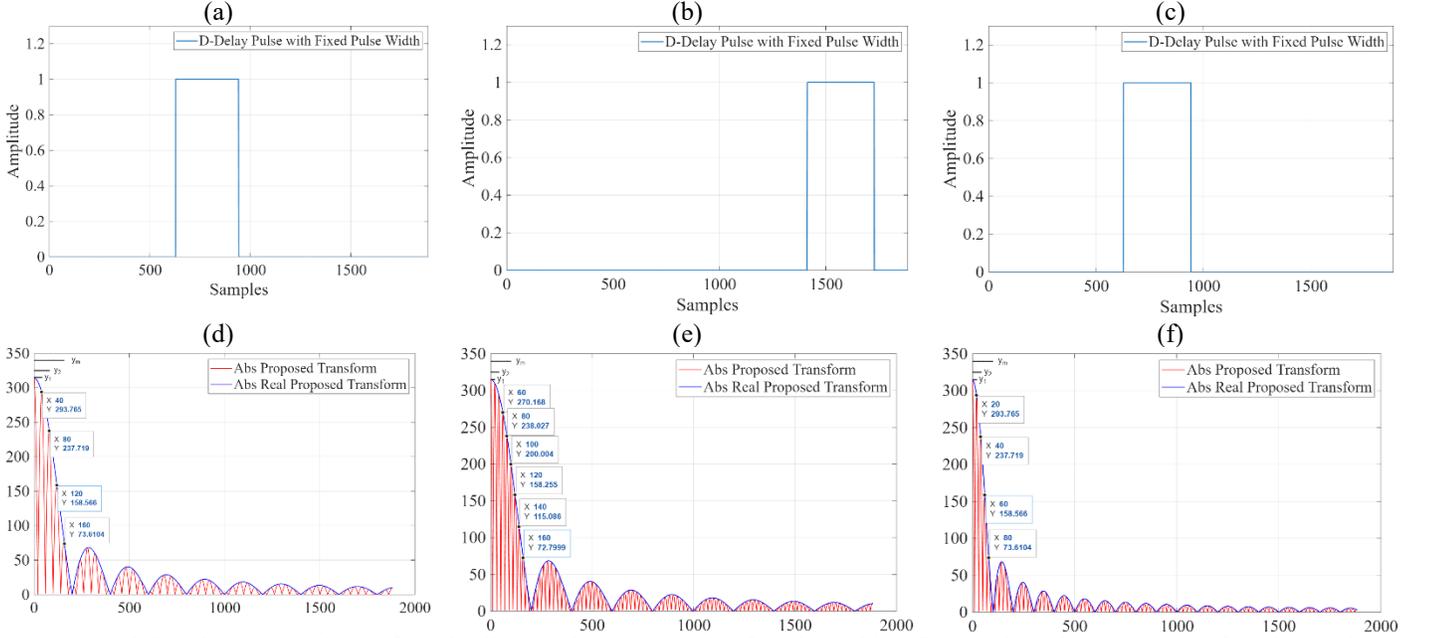

**Fig. 7.** Applying the proposed transform in discrete form on a pulse function with a fixed pulse width. (a) Pulse function with D-delay. (b) The proposed forward discrete transform of (a) with the transfer function $f_o$. (c) Pulse function with 2D-delay. (d) The proposed forward discrete transform of (c) with the transfer function $f_o$. (e) Pulse function with D-delay. (f) The proposed forward discrete transform of (a) with the transfer function $2f_o$.

As shown in Fig. 7 (d)-(f), the proposed transform provides a zoomed and precise frequency estimate. The absolute value of the real part of the proposed transform, Eq. (12), is shown in a blue dashed line, representing the envelope of the fringes generated by the absolute value of the real and imaginary parts. Thanks to the fringes generated by the absolute value of the real and imaginary parts of Eq. (12), the original signals can be easily reconstructed using the equation below:

$$y_m = \frac{m.h_0.N.f_s}{d.f_o}$$

(15)

Where $h_0$ is a constant, $y_m$ is the distance between fringes, d is the delay measured in samples ($\frac{d}{f_s}$ is the delay in seconds), $f_o$ is the transfer frequency, $f_s$ is frequency sampling, and N is the number of sample points. Since $e/f$ does not change for a scale-invariant observer, it guarantees that $h_0$ is a constant.

The fringes can help to recover the delay and transfer frequency used in the proposed transform. In addition, the pulse analysis allows us to understand how transfer frequency and delay affect the precision of frequency estimation. As shown in Fig. 7, the study of a rectangular pulse signal indicates that the transfer frequency should be smaller to achieve a higher frequency estimate. Moreover, the signal should be delayed to obtain even more fringes within the envelope, leading to a higher-frequency estimation. On the other hand, a lower transfer frequency eliminates high frequencies, so the reconstructed rectangular pulse cannot be well reconstructed.

If we consider signals as the sum of a DC component and the original signal, the DC component will represent the pulse signal and will not be reconstructed well. Hence, we need to remove the signal mean before using the proposed transform for better envelope extraction, and introducing a delay will lead to higher-frequency estimation and improved signal reconstruction. Thus, we introduce some delays into the FID signals and remove the signal mean in the experiments.

Although the DC component of the signal will be removed, the residual signal will still contain a rectangular push.
Therefore, we first extract the envelope from the absolute value of the output of Eq. (14), then extract the carrier signal. Then, we extract the carrier envelope using Eq. (13) and subtract it from the original envelope, leading to a finer envelope extraction, as will be seen in the following subsection.

### B. Synthetic Data Study

A synthetized noisy FID signal was constructed based on Eq. (4) with a Lamor frequency of 5 Hz, amplitude 10V, relaxation time ($T_2^*$) 250 $ms$, and the initial phase 0. The signal was sampled at a frequency of 100 Hz with a sampling interval $T_s$=(1/f s )= 0.01 s, resulting in 1885 data points over a time period of $6\pi\ s$.

In NMR signal processing, the observed signal is typically corrupted by a combination of different noise sources. The most common types include Rician-distributed noise, Gaussian (white) noise, Johnson (thermal) noise due to heat generation in the device's magnetic cores, and harmonic interference from the power line. To model an ideal noisy signal, we consider a mixture of these noise components with the following characteristics: Rician-distributed noise is simulated by adding random samples drawn from a noncentral chi distribution with a noncentrality parameter of 1, a scale parameter of 0.3, with approximately zero mean and amplitude of 7 $V$, Gaussian random noise with zero mean and a standard deviation of $0.15\ V$ and amplitude of $3\ V$, Johnson noise calculated for the highest temperature of the coils measured at the lab, which was $43\ ℃$, and harmonic interference simulated as a sinusoidal disturbance at $50\ Hz$ with amplitude $0.3\ V$. The constructed FID signal is shown in Fig. 8(a).

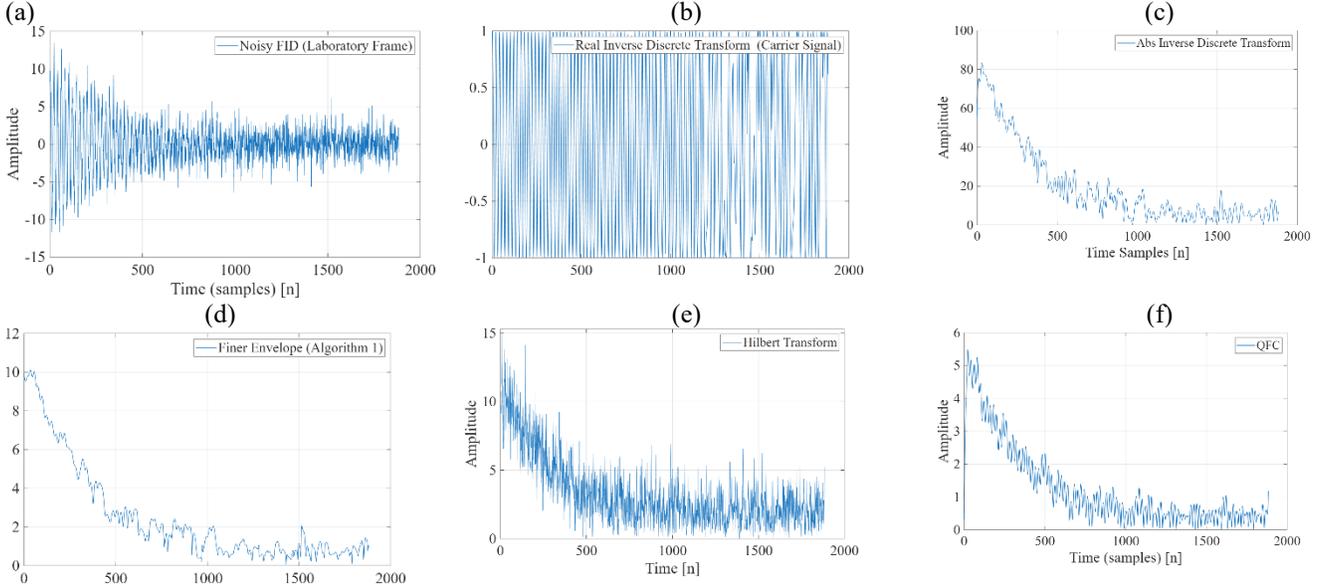

**Fig. 8.** Applying the proposed transform in discrete form on a synthesized NMR FID signal. (a) A synthesized NMR FID signal. (b) The extracted carrier of (a) using the inverse discrete transform. (c) The extracted envelope of (a) using the inverse discrete transform. (d) Applying the proposed inverse discrete transform on (c). (e) Extracted envelope by the Hilbert transform. (f) Extracted envelope using the QFC [3].

As discussed, there is no need to adhere to the Nyquist rate and the number of samples equal to the transfer frequency ($f_o$) is sufficient in the proposed transform. For this reason, we preprocess the original signal by downsampling it by a factor of 8 to be processed by the proposed transform and by the Hilbert transform; however, in other methods, in many cases, due to the use of Fourier transforms or LTI filters, it is necessary to meet the Nyquist rate. For example, in the QFC method, because band-pass filters are used, the Nyquist rate must be satisfied. This is the reason for the difference in the number of samples observed in Fig. 8. (c)-(f). Despite the different number of samples, the time is the same in all of them.

The proposed transform produces a finer envelope in Fig. 8(c) than the Hilbert transform and QFC methods, as seen in Fig. 8(e) and (f), respectively. In addition, the extracted envelope gets finer with reprocessing by the proposed transform, as shown in Fig. 8(d).

### C. Real NMR Data Study

The real FID and spin-echo signals used in this paper are obtained via the setup shown in Fig. 2. The Earth's magnetic field was measured in the laboratory to be $39.75\ \mu T$ with a magnetometer, whose Larmor frequency was calculated to be 1692 Hz. The polarization of the magnetic field $B_p$ is $25\ mT$. We used a sinusoidal RF pulse with a rectangular push with $1.2\ ms$ RF pulse duration. The acquisition time is set to 5s for the water sample. We did not use any averaging operation of the oscilloscope to acquire the NMR signals.

The spin-echo signal from a tube of water exposed to high-field homogeneity, shown in Fig. 3(b), is cropped and processed. The extracted envelope of the proposed transform is compared with those of the classic Hilbert transform and the QFC method. As shown in Fig. 9, the proposed transform extracts a finer envelope than the other methods; moreover, it analyzes a larger number of samples because it does not require meeting the Nyquist rate.

### V. Conclusion

We challenged the classic definition of scale, which primarily relies on physical properties of objects. We defined scale based on variations in space-invariant states, rather than on observable characteristics of matter and energy. According to the new perspective on scale, we confined the observer to three properties and categorized observers into two types: scale-variant and scale-invariant. These properties help to confine the frames to unique observers. Then, we introduced a novel transform that converts a scale-variant observer into a scale-invariant observer. We showed that the envelope of the NMR signal in the laboratory frame is scale-variant, whereas the signal in the rotating frame is scale-invariant. The introduced transform generates an analytic signal whose absolute value yields the envelope of the NMR signal, and this observation is scale-invariant. In addition, the advantage of the proposed transform is that it does not require meeting the Nyquist rate. The transform outperforms the classic and state-of-the-art methods for NMR signal envelope extraction.

### References

[1] D. Grombacher, L. Liu, M. A. Kass, G. Osterman, G. Fiandaca, E. Auken, and J. J. Larsen, "Inverting surface NMR

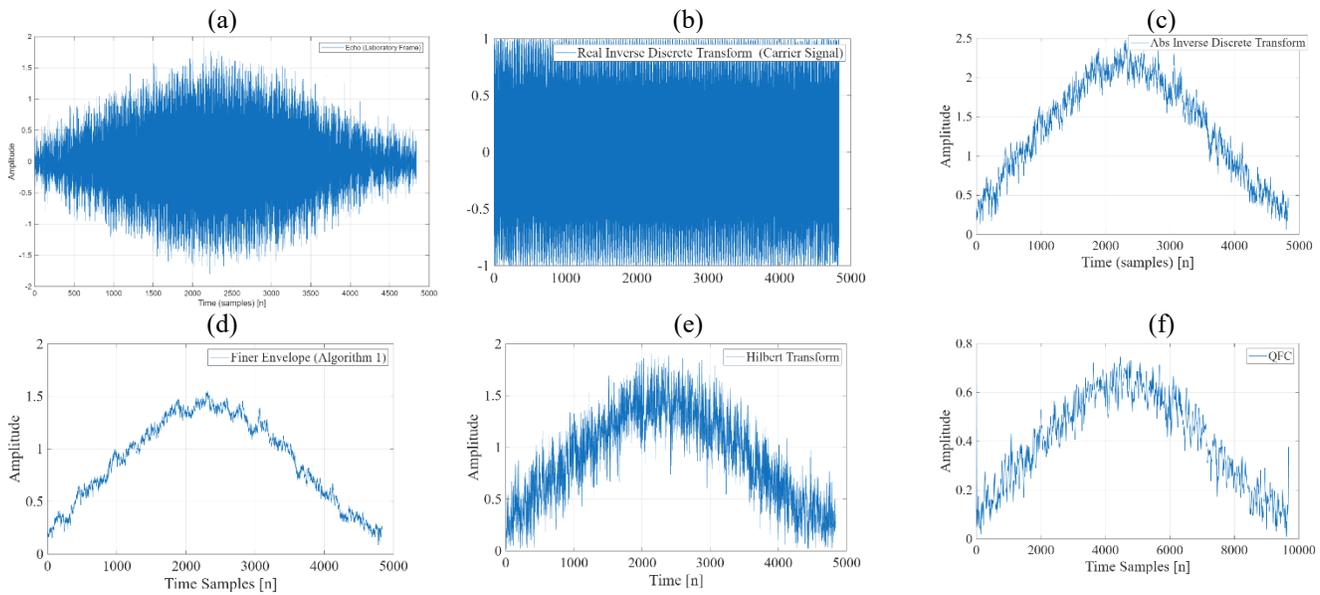

**Fig. 9.** Applying the proposed transform in discrete form on a real NMR spin-echo signal. (a) NMR spin-echo signal of a tube of water. (b) The extracted carrier of (a) using the inverse discrete transform. (c) The extracted envelope of (a) using the inverse discrete transform. (d) Applying the proposed inverse discrete transform on (c). (e) Extracted envelope by the Hilbert transform. (f) Extracted envelope using the QFC [3].

free induction decay data in a voltage-time data space," Journal of Applied Geophysics, vol. 172, p. 103869, Nov. 2020.

[2] L. Liu, D. Grombacher, M. P. Griffiths, M. Ø. Vang, and J. J. Larsen, "Signal processing steady-state surface NMR data," IEEE Transactions on Instrumentation and Measurement, vol. 72, pp. 1–13, 2023.

[3] L. Tian, Y. Zhang, H. Wang, Q. Chen, and M. Li, "Third-order cumulants with quadratic frequency coupling constraints for robust MRS signal analysis," IEEE Transactions on Biomedical Engineering, vol. 72, no. 4, pp. 1023–1035, Apr. 2025.

[4] A. Abragam, *The Principles of Nuclear Magnetism*. Oxford, U.K.: Oxford Univ. Press, 1961.

[5] Z.-P. Liang and P. C. Lauterbur, *Principles of Magnetic Resonance Imaging: A Signal Processing Perspective*. New York, NY, USA: IEEE Press/Wiley, 2000.

[6] P. L. Carson, H. N. Yeung, M. O'Donnell, and Y. Zhang, "Optimal sampling strategies for the measurement of spin–spin relaxation times," Med. Phys., vol. 15, no. 5, pp. 767–775, Sep.–Oct. 1988.

[7] M. A. Khan and S. A. Malik, "Real-time FPGA implementation of a 90° phase shifter based on Hilbert transform," IEEE Access, vol. 12, pp. 117345–117356, Oct. 2024.

[8] Gabor, Dennis. "Theory of communication. Part 1: The analysis of information." *Journal of the Institution of Electrical Engineers-part III: radio and communication engineering* 93, no. 26 (1946): 429-441.

[9] M. Feldman, "Hilbert transform in vibration analysis," *Mechanical Systems and Signal Processing*, vol. 25, no. 3, pp. 735–802, Apr. 2011. doi: 10.1016/j.ymssp.2010.07.018